# The advantages of the pentameral symmetry of the starfish

Liang Wu[a,1], Chengcheng Ji[a,1], Sishuo Wang[a], and Jianhao Lv[b]

[a] College of Biological Sciences, China Agricultural University, Beijing, 100094, China
[b] College of Science, China Agricultural University, Beijing, 100094, China
[1] Joint first authors.

Corresponding author
Liang Wu
College of Biological Sciences, China Agricultural University, Beijing, 100094, China
Tel: +86-10-62731071/+86-13581827546
Fax: +86-10-62731332
E-mail: wuliang8910@gmail.com

Chengcheng Ji
E-mail: jichengcheng@yahoo.com.cn
Sishuo Wang
E-mail: xibumumin9@126.com
Jianhao Lv
E-mail: haojianlv@gmail.com

## Abstract

Starfish typically show pentameral symmetry, and they are typically similar in shape to a pentagram. Although starfish can evolve and live with other numbers of arms, the dominant species always show pentameral symmetry. We used mathematical and physical methods to analyze the superiority of starfish with five arms in comparison with those with a different number of arms with respect to detection, turning over, autotomy and adherence. In this study, we determined that starfish with five arms, although slightly inferior to others in one or two aspects, exhibit the best performance when the four aforementioned factors are considered together. In addition, five-armed starfish perform best on autotomy, which is crucially important for starfish survival. This superiority contributes to the dominance of five-armed starfish in evolution, which is consistent with the practical situation. Nevertheless, we can see some flexibility in the number and conformation of arms. The analyses performed in our research will be of great help in unraveling the mysteries of dominant shapes and structures.

## Keywords

## 1. Introduction

The term "starfish" refers to Asteroidea of the Echinoderm phylum [1]. Starfish are primarily marine bottom-dwellers [2]. They have numerous tube feet on ambulacral grooves. When the starfish adhere to the seabed, these tube feet stretch out and densely cover the oral surface [3,4]. Nearly all starfish are carnivorous and usually feed on mussels, and they use their olfactory sense to detect the position of their food [5,6]. Starfish also exhibit a high regenerative ability [7,8]. Starfish usually show pentameral symmetry, and they are typically close in shape to a pentagram, which exhibits the golden ratio [9].

Starfish have an evolutionary history that spans five hundred million years, beginning in the Cambrian period, wherein the dominate species have always exhibited pentameral symmetry [10]. Starfish with five arms may evolve to have other numbers of arms. For instance, six-armed starfish and numerous-armed starfish, which usually has 10-15 arms, both exist [11]. From a developmental point of view, five-armed starfish can readily turn into new types of starfish with different numbers of arms [11,12]. It turns out that starfish have a tendency to grow into a pentagram shape from starfish with thin arms [10], which demonstrates that starfish with thin arms or with other arm numbers have not out-competed starfish with pentagram shapes in most ecological niches. Here, we analyzed the superiority of starfish with five arms over those with other arm numbers in their econiche.

We first considered the functions of arms to analyze the advantages of different arm numbers. The arms of a starfish are used to crawl along the sea bottom [13], force open the shells of bivalve mollusks [14], take care of young [15,16], detect food [5,6], turn over [13], adhere to the seabed when buffeted by sea waves [3,4] and escape via autotomy under emergency conditions [17]. We qualitatively analyzed the first three functions and quantitatively analyzed the detection of food, turning over, autotomy and adherence to the seabed.

Starfish take on a variety of forms that are expressed externally by a variety of structures, such as spines and granules [18,19]. Each different form has a different thickness; however, the two-dimensional graph of a pentagram is relatively conserved. This conservation demonstrates that the two-dimensional graph of a pentagram is the key to preserving the five-armed starfish's dominance in evolution. Because the pentagram exhibits the golden ratio, its evolutionary advantage has research value beyond the study of starfish. We concentrated on the advantages provided by having a two-dimensional graph of a pentagram shape.

We used mathematical modeling to explore the advantages of a two-dimensional pentagram graph. Based on the restriction of the two-dimensional shape of five-armed starfish and taking into consideration the actual two-dimensional shapes of five-armed and six-armed starfish, we eventually developed two-dimensional graphs for other arm numbers. We then used mathematical modeling to simulate the detection of food, turning over, autotomy and adherence to the seabed, and we determined the advantages and disadvantages of different arm numbers for the different functions. Finally, we integrated the performance of the four aspects that we tested to compare the benefits of different numbers of arms.

## 2. Methods

## 2.1 Modeling

When constructing the two-dimensional graphs of starfish with other numbers of arms, we first considered the most important factors that constrain the two-dimensional graph of a five-armed starfish. For living creatures, the most important attribute is always the ability to acquire energy. For starfish, both the arm and disc are preying devices, and the longer the arm of the starfish is, the better its ability to prey. Therefore, we restricted the radius of a starfish to a fixed value (set to $(\cot\frac{\pi}{5} + \cot\frac{\pi}{10})$ cm). After referencing actual five- and six-armed starfish in nature, we discovered that we should also fix the ratio of the area of the disc to the total area of the starfish, which equals to $\cot\frac{\pi}{5}/(\cot\frac{\pi}{5} + \cot\frac{\pi}{10})$ in five-armed starfish. We then developed different two-dimensional graphs for different numbers of arms, as shown in Fig. 1.

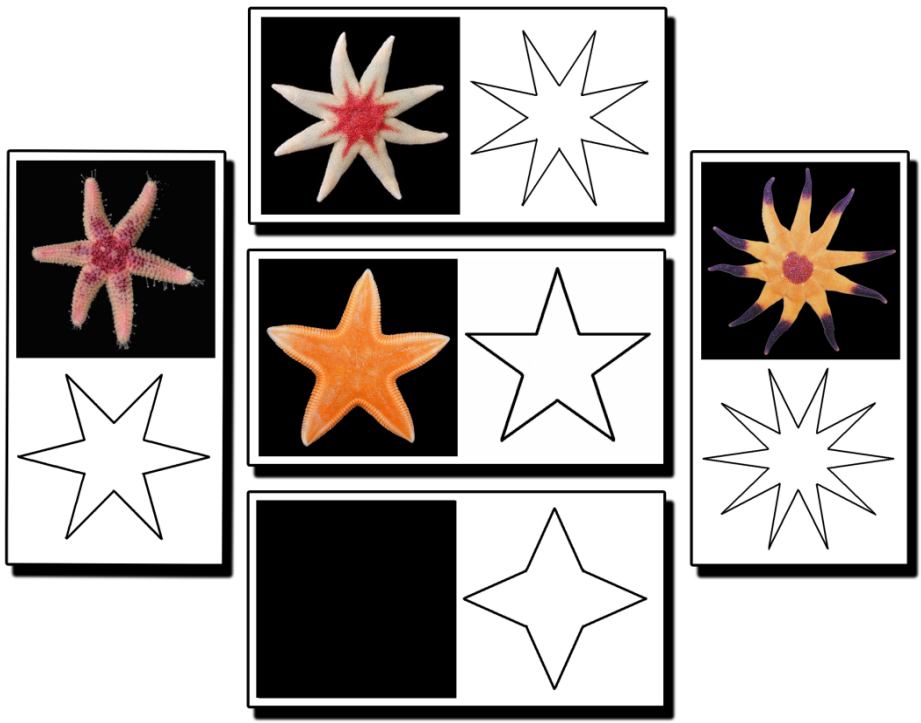

**Fig 1. Modeling.**
Real starfish compared to the output of the mathematical model. No four-armed starfish is shown in the picture, as no species of starfish is characterized by four arms [11].

As shown in Fig. 1, the model results are consistent with the actual examples. Here, we explored the advantages and disadvantages of four-, five- and six-armed starfish in comparison with one another. No starfish exist that have three arms, and we speculated that three-armed starfish do not have enough arms to survive. Numerous-armed starfish, such as sunstars, greatly differ in their living habits from five-armed starfish, eating almost anything, including smaller starfish and even smaller sunstars [20,21]. Numerous-armed starfish, therefore, might need more arms for better detection and faster locomotion, and we therefore excluded them from this study. Therefore, four-armed starfish, among starfish with fewer arms, are the starfish comparable to five-armed starfish. Six-armed starfish are also included, as they can be regarded as representative of starfish with more, but not numerous, arms.

## 2.2 Detection

Starfish use olfaction to detect food [5,6]. The odor molecules released by the food are received by the starfish, allowing the direction and position of the food to be sensed. The odor molecules are subject to Brownian motion. We used the random walk model to simulate Brownian motion, and the food is sensed if the wandering point reaches the target region. To ensure that the migration can reach the target region, we stipulated the following, the derivation of which is in the supplementary materials in section 1:

$$\sqrt{\text{Number of steps}} \times \text{step length} \approx \text{the distance from the the starting point to the center of the target region}$$

We limited the number of steps, which represents the amount of time spent in an actual situation, and we counted the number of wandering points that enter into the target region, which represents the number of odor molecules that reach the starfish, and the number of steps, which represents the time taken to reach the starfish. Because starfish and their food all exist on the plane of the seabed and because an odor molecule will only be detected if it returns to this plane, we can project this three-dimensional process onto a two-dimensional plane. The angle from the starting point to the target region will influence the results; hence, we obtained the average of all possible angles. We set the distance of the starting point from the center of the target region at 2, 3, 4, 6, 8, 10 and 15 times the radius of the starfish. (The detailed calculation process is included in the supplementary materials in section 1.)

## 2.3 Turning over

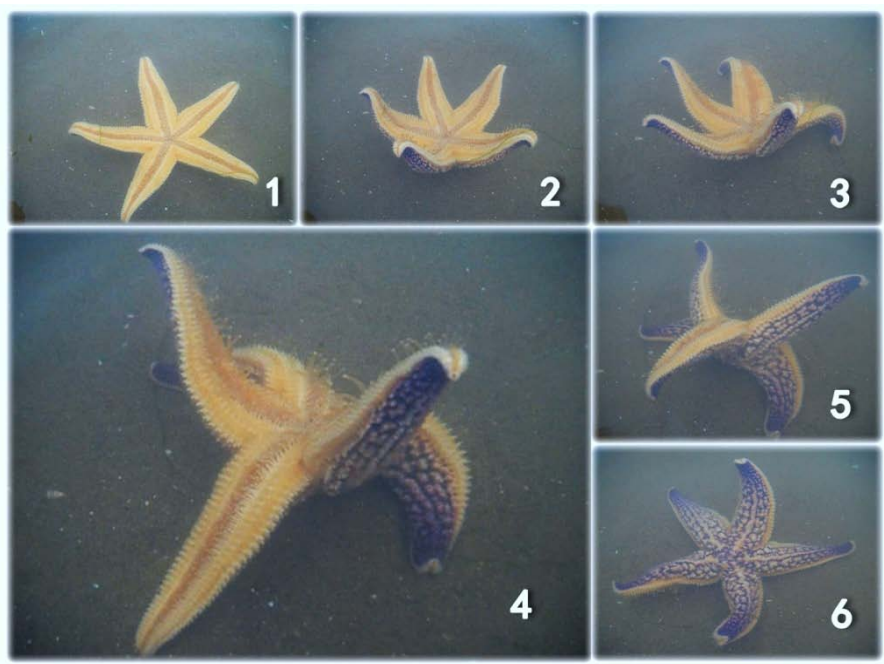

**Fig 2. Turning-over process.**
The turning-over process is shown from Step 1 to 6. Step 4, in which the center of gravity is passing the critical state, determines whether a starfish can successfully turn over.

We found that starfish living in the epicontinental seas are often turned over due to the motion of sea waves, exposing their oral surface and leaving them vulnerable to their natural enemies or to drying out. The ability to turn back over is crucial. From the procedure shown in Fig. 2, we can conclude that whether a starfish can successfully turn over is determined by whether the center of gravity can pass the edge of the portion of the supporting arms that are against the ground. From observing starfish, we can see that they try to keep their center of gravity as low as possible, which also supports the theory of lowest energy consumption. We simulated whether the center of gravity of a starfish can pass over the two supporting arms under conditions similar to those experienced by starfish in the wild (shown in Fig. 3). In the model, we approximated that two supporting arms bend into the shape of circular face, and the other three arms are perpendicular to the disc. We set the radiuses of the two supporting arms of the three types of starfish keep the same. The density of a starfish was based on *Asterias amurensis*. (The detailed calculation procedure is included in the supplementary materials in section 2.)

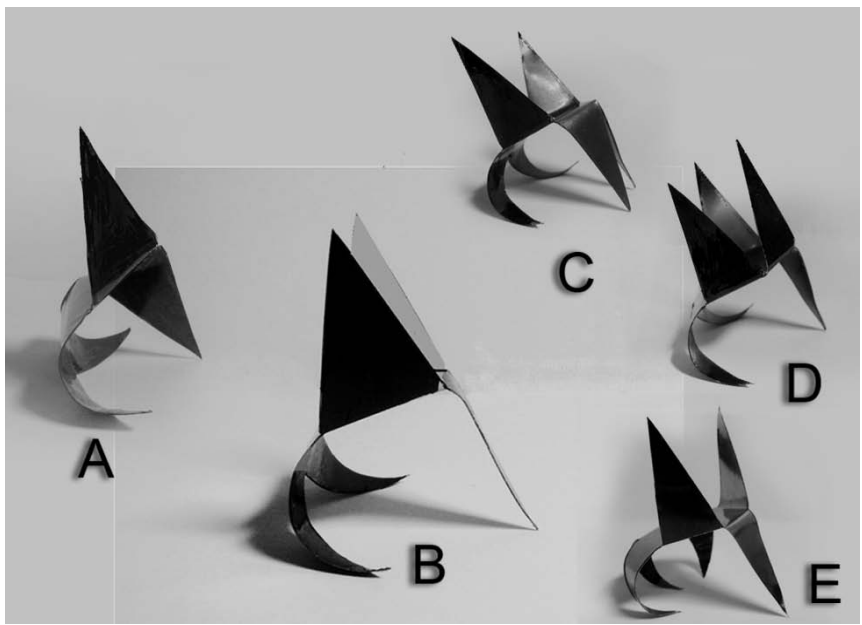

**Fig 3. Turning-over modeling.**
A and B are, respectively, the turning-over models for four-armed starfish and five-armed starfish. C, D and E represent the three possible turning-over situations for six-armed starfish.

## 2.4 Autotomy

Under sub-lethal situations, starfish will resort to autotomy. They will cut off one of their arms to escape and are capable of quickly regenerating another arm [7,8,17]. This indicates that the process of autotomy is a matter of life and death. Under proper conditions, the arm of some species that is cut off can also grow into a new starfish; thus, autotomy can also be seen as a form of asexual reproduction [22,23]. These two facts explain why the process of autotomy is crucial in starfish.

Pentagram starfish always lose their arms near the disc, whereas the autotomy of starfish with thinner arms can take place anywhere along the arm [7,8,17,23]. This difference suggests that autotomy near the disc is highly related to the starfish's overall conformation. We discussed only the pentagram form of autotomy. Unlike other types of mechanical stimuli, cutting the tube feet of a specific arm will cause the instant autotomy of that arm [8,24,25]. In our experiment, we found that an upward force is more likely to fracture the tube feet than a directional lateral force. In nature, an upward force would come from a predator's pulling motion. This observation supports the fact that autotomy primarily occurs under the threat of predation [21,26,27,28,29]. During autotomy, starfish will release an autotomy-promoting factor to act on the proximal region of the arm and result in an endogenous loss of tensile strength in the connective tissues at the autotomy plane [30]. Because this chemical action is the same for the three types of starfish, we did not include this factor in our analysis. We used the mechanics of the materials to analyze the process of autotomy, and our conclusions are demonstrated by the force diagram in Fig. 4.

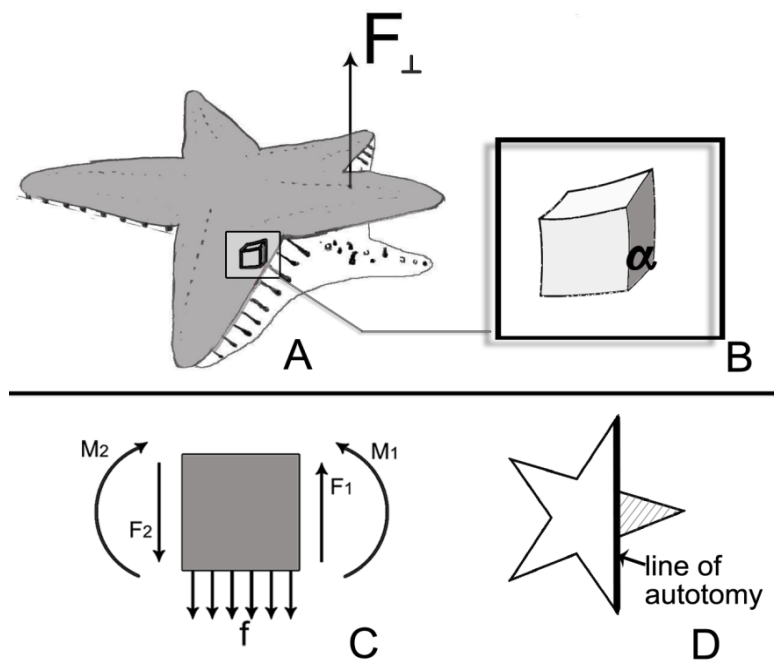

**Fig 4. Schematic diagram of autotomy.**
A. An outside force acts on the starfish.
B. The shape of the body wall changes.
C. Each small piece of the body wall is affected by the bending moment (M1, M2), stress (F1, F2) and the resisting force (f) of the tube feet (seen from the angle of α face).
D. The line of autotomy, indicated by the arrow, has a pivotal role.

As shown in Fig. 4, the upward pulling force applied by a predator attacking a starfish acts on the arm, resulting in a curling of the body wall, which, in turn, elongates the tube feet to provide a resisting force against the upward pulling force. The resisting force is in equilibrium with the pulling force, which allows it to end with a fracture at the proximal region of the arm. In this process, the tube feet on other parts of the starfish body are not involved. Only the number of tube feet around the black line determines the likelihood of autotomy. The more tube feet there are, the larger the resisting force, and the higher the probability that it will end in equilibrium with the pulling force. Therefore, the possibility of autotomy is positively correlated with the number of tube feet. Because the tube feet on ambulacral grooves stretch out when adhering, the oral surface is densely covered with tube feet along the black line in equal density. The longer the line, the more tube feet there will be, and the larger the resisting force they can produce. We named this line the **line of autotomy**. We simply need to compare the length of this line between four-armed, five-armed and six-armed starfish to understand whose resisting force is larger. When the outside force is set, a longer line of autotomy increases the amount of force it can produce during autotomy. This increased force, in turn, increases the possibility of balancing the outside force and results in a higher probability of autotomy. Therefore, autotomy is more likely to occur with a longer line of autotomy.

## 2.5 Adherence

After close observation, we found that the only force that can resist the impact of sea waves is the adherent force to the seabed [3,4]. When the water velocity is fixed, the impact force is lessened, or the adherent force to the seabed is increased, a starfish is more likely to resist the wave's impact forces, increasing its odds of survival. Therefore, we defined the adherent effect as follows:

$$\text{Adherent effect} = \text{adherent force} / \text{impact force}$$

Adherent force is proportional to the area of tube feet, and we set the area of tube feet / total area to be a constant value. By combining these equations, we got:

$$\text{Adherent effect} \propto \text{area of the starfish} / \text{impact force}$$

By calculating the impact forces (obtained using fluid dynamics with the help of the Fluent software) for the three types of starfish, subjected to an identical current and with identical starfish area, we can determine the adherent effect quality of each starfish type. Because the direction of the current will affect the impact force, we took the average of all angles of current flow. During field research, we noted that the velocity of flow often reached 2 m/s; hence, we set the current velocity to 2 m/s. We set the density and viscosity to the values for seawater.

# 3. Results

## 3.1 Detection

Using the random walk model, we measured the number of odor molecules that reach four-armed, five-armed and six-armed starfish over a range of distances. We used the case of a five-armed starfish as the control and set the percentage of molecules that reach each five-armed starfish to 100%. By normalizing the remaining data, we were able to produce the data in Fig. 5.

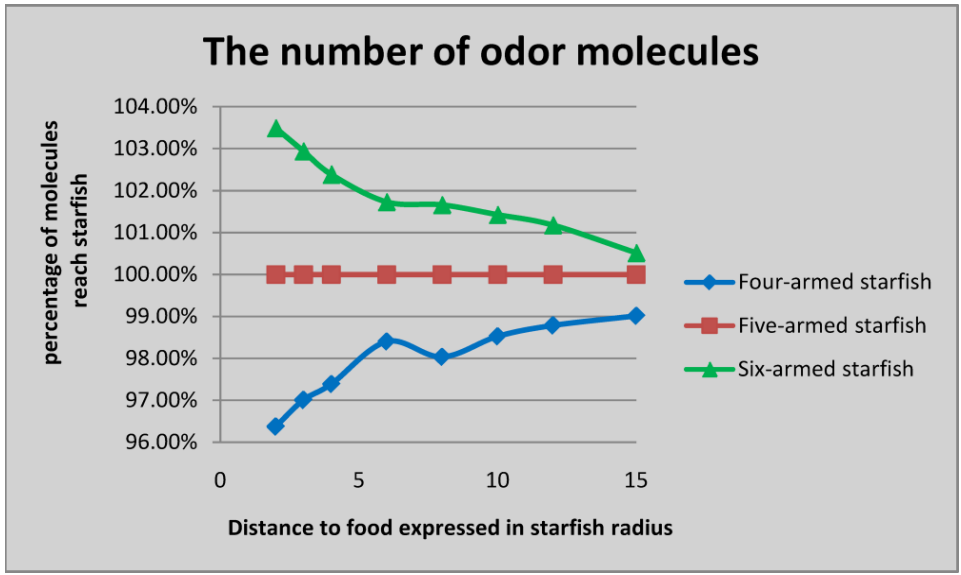

**Fig 5. The number of odor molecules.**
The x-axis represents the distance of the starting point from the center of the target region at 2, 3, 4, 6, 8, 10 and 15 times the radius of the starfish. The y-axis represents the percentage of molecules calculated to reach each starfish (regarding that of five-armed starfish as 100%). We can see that the number of odor molecules reaching six-armed starfish is greater than the number reaching five-armed starfish, which in turn is more than for four-armed starfish.

As previously mentioned, the number of steps that the migrating spot took to reach the target region represents the time required for the odor molecules to reach the starfish. We set the number of steps required for the spot to reach the target region of a five-armed starfish to 100% and normalized the remaining tests to produce the data in Fig. 6.

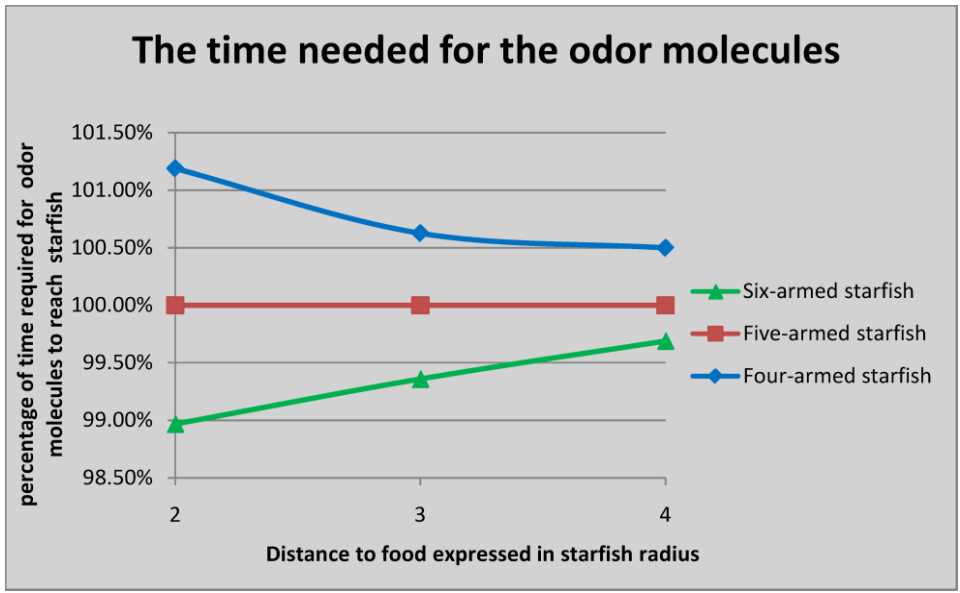

**Fig 6. The time needed for the travel of odor molecules.**
The x-axis represents the distance of the starting point from the center of the target region at 2, 3, and 4 times the radius of the starfish. The y-axis represents the percentage of time required for the odor molecules to reach the starfish (regarding that of five-armed starfish as 100%). The time six-armed starfish require is less than five-armed starfish, which in turn require less time than four-armed starfish.

As can be seen from Fig. 5 and Fig. 6, a higher number of arms results in better food detection, as one might expect. From a biological viewpoint, arms receive the olfactory information. From the viewpoint of this mathematical model, a larger number of arms increases the perimeter distance, which, in turn, allows the wandering point to more easily touch an edge of the target region.

From Fig. 5 and Fig. 6, we can see that under identical circumstances, the number of odor molecules that a six-armed starfish detects is greater than that of a five-armed starfish, which, in turn, is greater than what a four-armed starfish would detect. When the distance to food is twice the radius of the starfish, the gradient difference is approximately 3.5%. The time required to detect food shows a similar tendency; that is, a six-armed starfish takes less time to detect food than a five-armed starfish, which, in turn, takes less time than a four-armed starfish．When the distance is twice the radius of the starfish, the gradient difference is about 1.1%.

Because it is advantageous to receive more odor molecules faster, we concluded that six-armed starfish outperform five-armed starfish in the detection of food. In turn, five-armed starfish are better at detecting food than four-armed starfish.

## 3.2 Turning over

As mentioned in the methods, the relative position of the center of gravity and the edge of the portion of the supporting arms against the ground determines the difficulty level of turning over. The farther the center of gravity is raised over the edge of contact, the more easily the starfish can turn over. We used the angle in Fig. 7 to scale the extent of deviation. The larger the angle is, the more easily the starfish will turn over. For a four-armed starfish, the angle is 104.68°, whereas the angle for a five-armed starfish is 89.65°, and a six-armed starfish has potential angles of 82.89°, 84.58° and 69.55°.

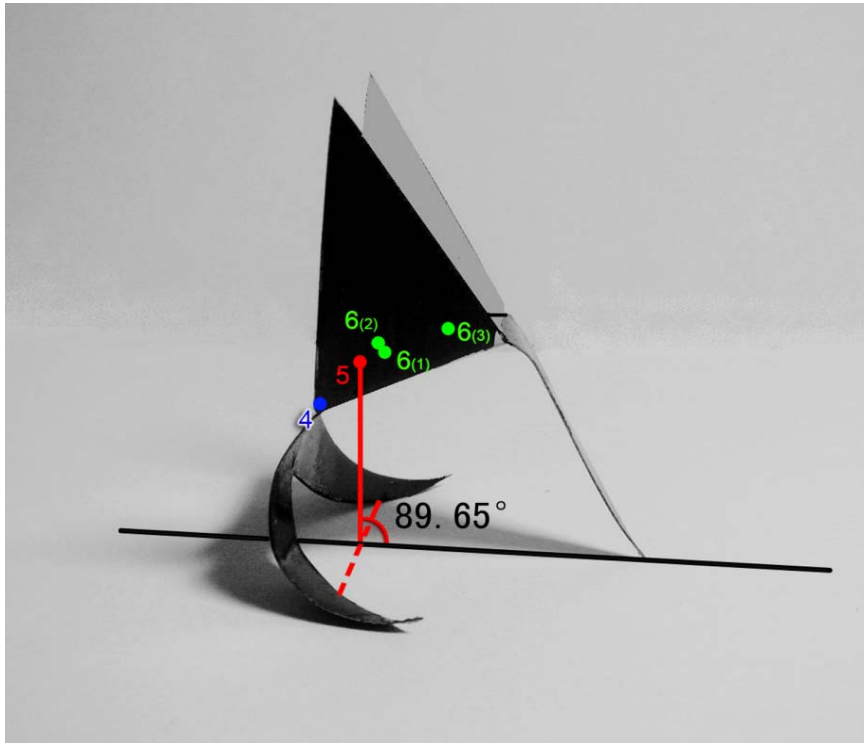

**Fig 7. Results of turning over.**
The blue spot is the center of gravity for a four-armed starfish. The red spot is the center of gravity for a five-armed starfish. The three green spots represent the center of gravity for a six-armed starfish under three different possible turning-over situations. The dotted line indicates the edge of the portion of the supporting arms that are against the ground; thus, the angle for a five-armed starfish is 89.65°. The degrees of the angles and the models of four- and six-armed starfish are not shown.

We can conclude that a four-armed starfish is more likely to turn over successfully than a five-armed starfish, which, in turn, is more likely to turn over than a six-armed starfish (we also calculated the angles required for seven- and eight-armed starfish to turn over; these are attached in the supplementary materials in section 2).

## 3.3 Autotomy

As mentioned in the methods section, the tube feet provide resisting force, and a longer line of autotomy corresponds to higher amount of force produced by the tube feet during autotomy. This increased potential force increases the chances of balancing the outside force, which results in a larger chance of autotomy. Therefore, it is better to perform autotomy with a longer line of autotomy. The blue lines indicated in Fig. 8 represent the **lines of autotomy** for the three types of starfish. In the four-armed starfish, only the tube feet of the arm that is being cut produce a resistant force during autotomy, and the tube feet of the adjacent arms are not involved. This situation results in a line of autotomy that is 2.7528 cm in length. Meanwhile, the line of autotomy is 8.4721 cm for five-armed starfish. In four-armed starfish, there is not enough resisting force generated to complete autotomy, leaving four-armed starfish vulnerable to predation. The line of autotomy for six-armed starfish is 7.9465 cm, demonstrating that a five-armed starfish also outperforms a six-armed starfish with respect to autotomy.

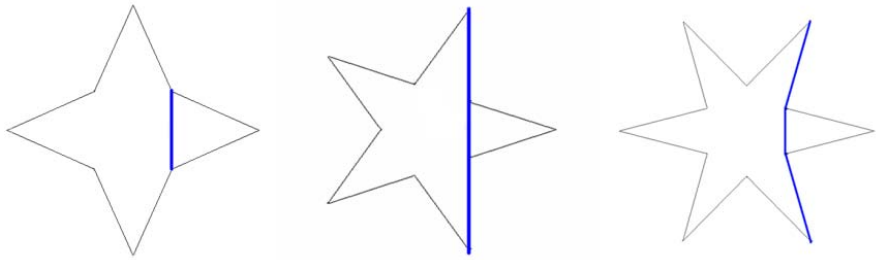

**Fig 8. Result of autotomy.**
The line of autotomy is marked in blue. The cases for four-armed, five-armed and six-armed starfish are shown. The line for four-armed starfish is much shorter because the tube feet of the adjacent arms are not involved in autotomy.

We concluded that five-armed starfish perform best at autotomy, whereas four-armed starfish are the least likely to perform autotomy.

## 3.4 Adherence

We calculated the impact force and the area of each type of starfish, and by using the formula 'adherent effect $\propto$ area of the starfish / impact force', we derived the results shown in Table 1.

Table 1 Adherent effect

|  | Area of the starfish (cm$^2$) | Impact force (N) | Area of the starfish /Impact force (cm$^2$/N) |
| --- | --- | --- | --- |
| Four-armed | 24.522 | 586.23 | 0.041830 |
| Five-armed | 22.270 | 600.15 | 0.037108 |
| Six-armed | 21.237 | 584.67 | 0.036322 |

Footnote: The impact forces are obtained using fluid dynamics with the help of the Fluent software. The model and initial data are included in the supplementary materials in section 3.

Regarding adherence, the four-armed starfish is better than the five-armed starfish, which is, in turn, better than the six-armed starfish. Under the same applied impact current, a four-armed starfish is more likely to adhere to the bottom of the sea than a five-armed starfish, which is in turn more likely to adhere than a six-armed starfish.

# 4. Conclusions

Our measurements indicate that each type of starfish has its own advantages and disadvantages in terms of the four examined attributes. To draw a final conclusion as to the optimal number of arms, we need to integrate the results from these four tests. We designated the five-armed starfish as the standard and set it to 1. After integrating the number of odor molecules detected from food at a distance of 2 times the starfish's radius, the turnover angles (we used the optimal angle for the six-armed starfish), the length of the line of autotomy and the adherent effect, we produced the results shown in Fig. 9 and Fig. 10.

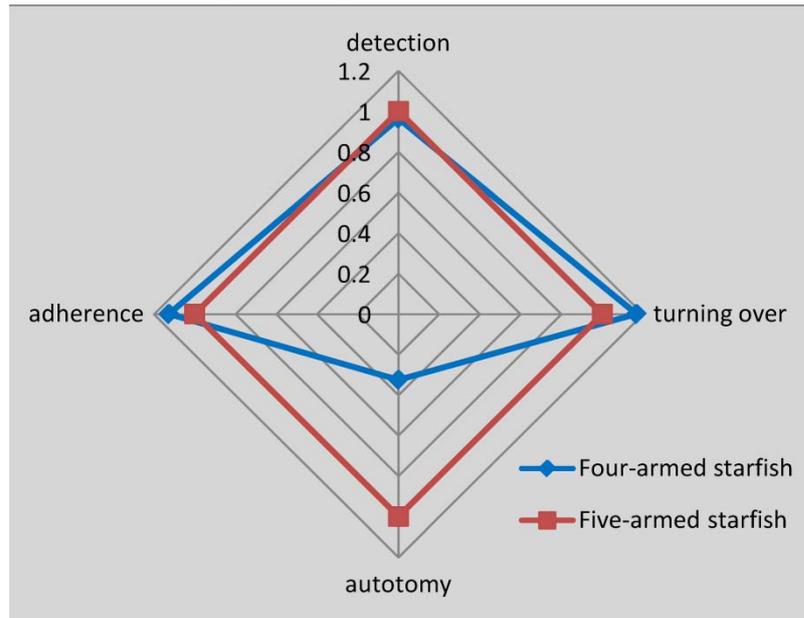

**Fig 9. Comparison between four-armed starfish and five-armed starfish.**

The four-armed starfish is shown in blue, and the five-armed starfish is shown in red. In comparison to four-armed starfish, five-armed starfish are similar in terms of detection, turning over and adherence, but they have a significant advantage in terms of autotomy. We can see from the figure that five-armed starfish have more advantages than four-armed starfish.

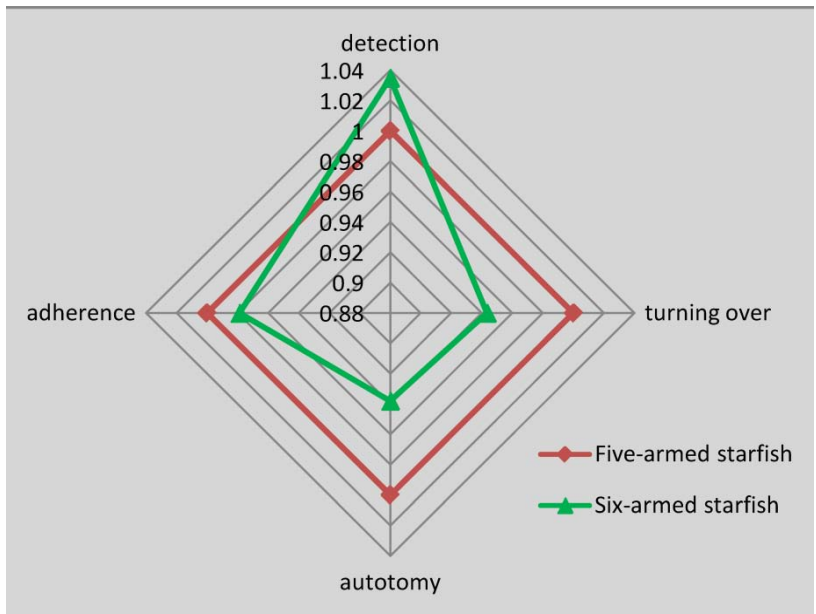

**Fig 10. Comparison between five-armed starfish and six-armed starfish.**
The five-armed starfish is shown in red, and the six-armed starfish is shown in green. In comparison to six-armed starfish, five-armed starfish have advantages in turning over, autotomy and adherence. Six-armed starfish are only superior in food detection. We can see from the figure that five-armed starfish perform better than six-armed starfish.

We can draw the following conclusions:
1. In comparison to four-armed starfish, five-armed starfish are similar in terms of detection, turning over and adherence but have a significant advantage, 208%, in terms of autotomy, leading us to conclude that, in general, five-armed starfish have an advantage over four-armed starfish.
2. In comparison to six-armed starfish, five-armed starfish have a 3.5% advantage in turning over, a 6.2% advantage in autotomy and a 2.13% advantage in adherence. Six-armed starfish are only superior in food detection, by 3.5%. This combination of factors leads us to conclude that, in general, five-armed starfish perform better than six-armed starfish.

Of the four investigated aspects, autotomy is the most crucial for a starfish's survival, as mentioned above. A five-armed starfish performs much better than the other two in this important process and performs as well or better in the other three aspects. This performance provides a five-armed starfish a much more significant advantage than is depicted in the radar picture.

A behavioral advantage can easily provide a survival advantage in natural selection, and an accumulation of survival advantages can out-compete inferior species after sufficient generations [31,32]. Based on evolutionary theory and the statistics obtained in our experiment, we would speculate that five-armed starfish will tend to out-compete the other two types of starfish and become the dominant species. This speculation is consistent with the fact that, in nature, five-armed starfish hold an absolute advantage over starfish with other numbers of arms, as evidenced by the fact that six-armed starfish exist in only small numbers and by the fact that four-armed species are never found [11].

## 5. Discussion

In this article, we focused on pentagram-shaped starfish and examined the advantages and disadvantages of three different types of starfish with regard to food detection, turning over, autotomy and adherence. We found that each body plan has advantages; however, after integrating the four aspects, we concluded that five-armed starfish have more advantages than six-armed starfish in most respects and are equivalent to four-armed starfish in all other aspects but are superior in autotomy. Five-armed starfish perform the best with general regard to the four factors we tested, especially autotomy.

We noticed that there is significant practical evidence for the superiority of five-armed starfish with regard to autotomy. In the rare six-armed starfish, such as *Leptasterias hexactis*, their method of autotomy typically does not take place at the proximal end of the arm. Instead, a portion of the arm is more likely to be cut off [33]. In starfish with thin arms, autotomy can take place anywhere along the arm [7,8,17,23]. The pentagram two-dimensional graph demonstrates the occurrence of autotomy at the proximal end of the arm. Autotomy at the proximal end of the arm should simplify wound healing and regeneration [8,25,34,35]. This possible advantage in healing may be why starfish with thin arms evolved to have the shape of a pentagram. Regarding turning over, we noticed that a six-armed starfish (a mutation of *Asterias amurensis*) takes longer to turn over than a five-armed starfish, which is consistent with the results obtained by our model. The line of autotomy is straight only when the five-armed starfish is in the shape of a pentagram that exhibits the golden ratio. We hypothesized that a straight line conveys the force with the fastest speed and contributes to faster autotomy. The golden ratio exists widely in the biological world [9], and the results of this research may help to address unresolved questions regarding the golden ratio in other creatures.

To address movement, preying and the protection of offspring, we performed some qualitative analysis and predictions. When crawling, the starfish's arms are used as locomotive organs. More arms may contribute to better crawling, which can be demonstrated using a mechanical model. When preying, starfish use their arms to catch shells and snap them open; hence, more arms may be advantageous. In terms of protecting offspring, starfish use their body as a shelter, and more arms may be of great help. In the three aforementioned aspects, additional arms are advantageous in all cases, although additional arms also mean that more energy is consumed. There must be a balance between the advantages of additional arms and energy consumption to allow the starfish to optimally accomplish these three tasks; five arms may well be the balance point. As for numerous-armed starfish, such as Solasteridae, their habits significantly differ from those of five-armed starfish, including an enlargement in their potential food sources to include other kinds of starfish and their own kind [20,21], which requires better detection and faster locomotion. These needs may directly result in an increased number of arms.

A survival advantage as a result of behavioral advantages can result in one species out-competing an inferior species after a number of generations [31,32]. Based on the statistics obtained in our analyses, if four-, five- and six-armed starfish all exist in a specific genus, the advantages of five arms will permit five-armed starfish to evolve into the dominant species. Four-armed starfish exhibit such disadvantages that it is hard for them to survive, whereas the relatively smaller disadvantages (relative to five-armed starfish) of six-armed starfish might allow a few species to exist. In fact, six-armed starfish occasionally exist, while four-armed species are never seen in nature [11]. Starfish with fewer than four arms cannot survive, whereas six-armed starfish are representative of starfish with more than five arms but are not numerous.

We can conclude that the optimal number of arms varies under different conditions based on the results for the four aspects studied here. Six-armed starfish would be better adapted to an environment with the following conditions: 1. a slow current, which reduces the need for adherence and turning over; 2. low predation, reducing the need for rapid and robust autotomy; and 3. food scarcity. For example, *Leptasterias hexactis*, one type of six-armed starfish, live in an environment where autotomy is not very important [33]. The two-dimensional outline of a five-armed starfish is not restricted to a pentagram shape. For example, if the environment requires an excellent ability to move, then a starfish with long arms will have an advantage. Starfish with smoother and rounder body conformations, such as Asterinidae, may have sacrificed some of their ability to detect, turn over and autotomize to increase their ability to adhere to the seabed. (Supporting data is in the supplementary materials in section 3.)

In this study, we used mathematics to analyze the advantages for starfish of having five arms over other numbers of arms. Our analyses may be useful for other studies investigating conformational advantages. The random walk model discussed herein can be used for other osmotic creatures, and the fluid mechanics method can also be adapted for use in other aquatic creatures. The integration of the advantages and disadvantages of different aspects is very helpful to developing an understanding of the evolutionary advantages of specific biological traits. For instance, the advantages of five digits in most mammals could also be analyzed by using mathematics and physics to model grabbing and crawling functions. Modeling the visual stimulation of flowers with different petal numbers to insects carrying out pollination could help address the question of why angiosperms typically have five petals in their flowers.

## Acknowledgments

This work is supported by the Undergraduate Innovation Program of China Agricultural University. We thank Professor Baoqing Wang for his kind support and our friend Zhuo Sun for providing language help. We also thank Yuping Chen and Zitao Ma for their assistance in figure-editing. The authors also extend their heartfelt thanks to the reviewers for their valuable suggestions.